\documentclass{vgtc}                          

\ifpdf
  \pdfoutput=1\relax                   
  \usepackage{graphicx}                
  \DeclareGraphicsExtensions{.pdf,.png,.jpg,.jpeg} 
\else
  \usepackage{graphicx}                
  \DeclareGraphicsExtensions{.eps}     
\fi%

\graphicspath{{figures/}{pictures/}{images/}{./}} 

\usepackage{microtype}                 
\PassOptionsToPackage{warn}{textcomp}  
\usepackage{textcomp}                  
\usepackage{mathptmx}                  
\usepackage{times}                     
\usepackage{cite}                      
\usepackage{tabu}                      
\usepackage{booktabs}                  

\onlineid{0}

\vgtccategory{Research}

\title{Exploring Affordances for AR in Laparoscopy}

\author{Matheus Negrão\thanks{e-mail: mdnegrao@inf.ufrgs.br}\\ %
        \scriptsize INF - UFRGS %
\and Joaquim Jorge\thanks{e-mail: jorgej@tecnico.ulisboa.pt}\\ %
\scriptsize IST - ULisboa, INESC-ID %
\and João Vissoci\thanks{e-mail: jnv4@duke.edu}\\ %
 \scriptsize Duke University%
\and Regis Kopper\thanks{e-mail: kopper@uncg.edu}\\ %
 \scriptsize UNC Greensboro %
\and Anderson Maciel\thanks{e-mail: anderson.maciel@tecnico.ulisboa.pt}\\ %
     \scriptsize INF - UFRGS \\
     \scriptsize IST - ULisboa, INESC-ID
}
\teaser{
  \centering
  \includegraphics[width=0.8\linewidth]{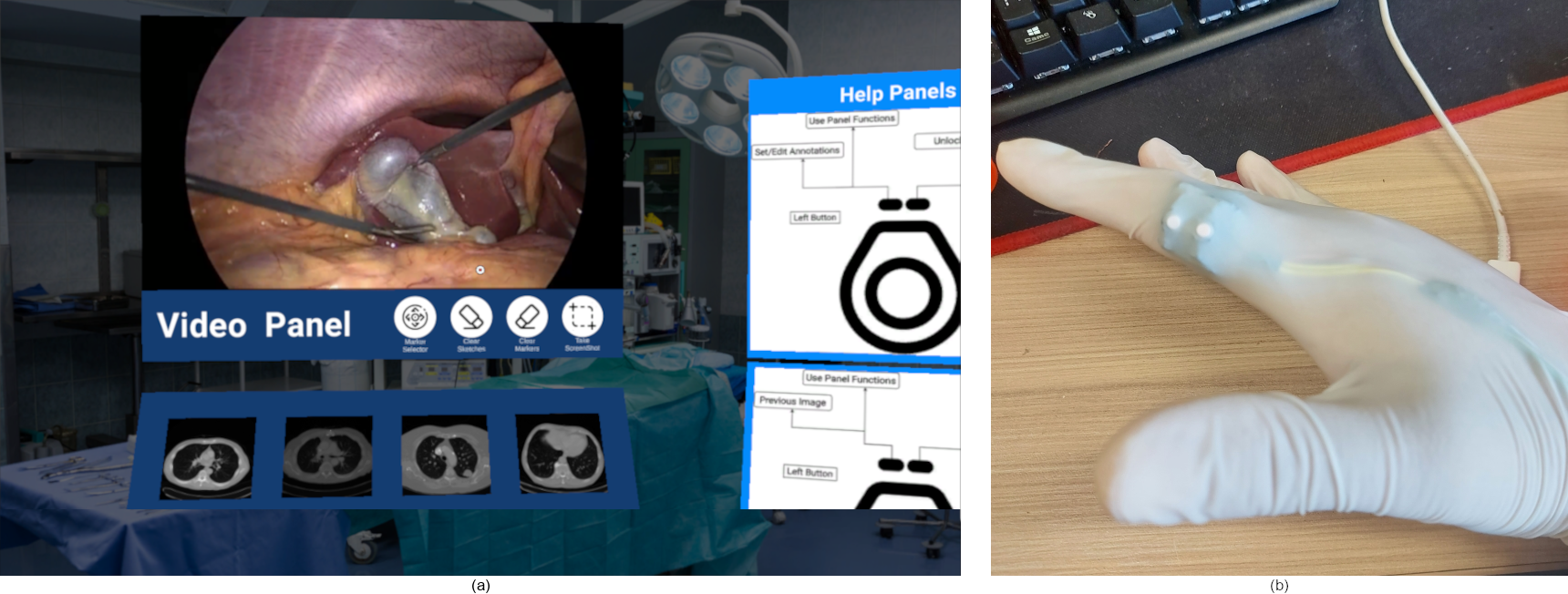}
  \caption{(a) The main view of the interface, with video, image, and help panels. (b) The ring device with two buttons for interaction.}
  \label{fig:teaser}
}

\abstract{
This paper explores the possibilities of designing AR interfaces to be used during laparoscopy surgery. It suggests that the laparoscopic video be displayed on AR headsets and that surgeons can consult preoperative image data on that display. Interaction with these elements is necessary, and no patterns exist to design them. Thus the paper proposes a head-gaze and clicker approach that is effective and minimalist. Finally, a prototype is presented, and an evaluation protocol is briefly discussed. 
} 

\CCScatlist{
  \CCScatTwelve{Human-centered computing}{Human computer interaction (HCI)}{nteraction paradigms}{Mixed / augmented reality};
  \CCScatTwelve{Human-centered computing}{Human computer interaction (HCI)}{nteraction paradigms}{Collaborative interaction}
}

\begin{document}


\firstsection{Introduction}

\maketitle

Minimally Invasive Abdominal Surgery (MIAS) has seen exponential growth and worldwide adoption. Due to less tissue trauma and lower pain scores by the patients, it allows for faster recovery after surgery~\cite{buia2015laparoscopic}. However, even seasoned surgeons face a variety of difficulties, namely: a) the lack of tactile feedback; b) the laparoscopic camera often held by less-experienced surgeons may result in deviations and rotations away from the target horizon, creating delays and loss of concentration. The surgical team is not looking at the intervention target as with open surgery, but instead, at monitors located 2 to 3 meters away. The latter, positioned at different angles/distances for each team member, requires an extra conscious effort to correct body posture. Furthermore, this setup limits access to preoperative images and planning data during surgery. When these data need to be consulted, the surgeon cannot easily leave the clean operation area and thus requests help from assistants who hold and manipulate these materials following verbal instructions, which is far from ideal. Finally, and more importantly, communication between members of the surgical team can be difficult and awkward. Since team members are holding instruments and executing complex tasks, it is difficult to point to "areas of interest" and disambiguate pointing references without losing time and concentration. This introduces pauses and hinders team communication and synchronization, increasing the potential for lost time and mistakes. Also, an operating room's complex environment and limited space around the table make it problematic to enlist specialists' help or instruct students, who could be located several feet away on the non-sterile field or farther away in different hospitals/countries.


AR and VR devices have become accessible, lightweight, and powerful while improving in quality and resolution, making them increasingly suitable as portable, high-definition, hands-free, and non-intrusive for usage in surgical settings. AR applications have been developed and evaluated, demonstrating effectiveness in laparoscopic skills training \cite{barsom2016systematic}. Some applications use AR head-mounted displays (HMD) to place objects in the real world and the surgical video, helping explain the training and execution of procedures. These systems can guide and identify the points of interest during the operation through shared appointments \cite{heinrich2021holopointer}, complementing verbal explanations and helping to reduce the complexity of the learning. In other cases, the application combines the views of surgery, vitals, and preoperative exams in the field of view to minimize the interruption and help with focus during the procedure \cite{al2020effectiveness}.

However, although AR systems can help in the training of laparoscopic surgeries and have been demonstrated in see-inside-the-body applications, intraoperative AR systems require interaction. They could not succeed due to a lack of understanding of how to interact with visual data during surgery effectively. The Microsoft HoloLens (HL) is commonly used for developing and evaluating augmented reality applications in related works. Some of these applications use voice commands and hand gestures as their primary method of interaction, both supported by HoloLens. However, using voice commands may be less accurate, and it may become confusing during procedures, as well as hand gestures, which make the surgeon pause the procedure, and take off their hands from the surgical instruments to interact with the interface and objects. A narrow field of view is an additional challenge. 

Therefore, in this paper, we explore the design space for interaction in intraoperative AR for laparoscopy. We do so throughout the development of a demonstrator that improves ergonomic conditions (e.g., using AR goggles as screens) in the operating room. Besides displaying the laparoscopic video on a virtual screen wherever the surgeon prefers to place it, the system provides access to preoperative images and planning data. The demonstrator also allows for real-time deictic communication through annotations, icons, drawings and other visual indications that may help with collaborative tasks~\cite{Sarmiento2013, Sarmiento2014}. The surgeons themselves can make such annotations and direct trainees and assistants to look at the important spots. 

Besides the novel interface design, our results include the planning of an evaluation of different interaction conditions that will be conducted as future work.

\section{Related Work}

Augmented and mixed reality have been applied to different tasks in the surgical environment, mainly using AR with head-mounted displays, such as Microsoft Hololens, for learning anatomy~\cite{Souza2020}, training procedures, surgery planning, and displaying information during surgery. The planning surgeries applications and prototypes explore the visualization of 3D reconstructions of organs in holograms, which the surgeons can have, based on preoperative exams, better visualization of the patient anatomy, structures, depth, and planning the intervention~\cite{morales2018prototype, sanchez2021application}. In laparoscopic surgeries, all the interventions are guided and visualized through the camera inside the patient, turning this video feed into the most crucial point in an application.


The HoloPointer \cite{heinrich2021holopointer} uses the head-gaze direction from Hololens to provide a local shared reference pointing in the 2D operation video monitor triggered by voice commands. All the staff in the operation room can view the exact point that the surgeon wants to show on the monitor, reducing errors and resulting in an economy of movement. In another work, the eye gaze is tracked and used for free-hand drawing and pointing in laparoscopy video to guide novice surgeons through a standard 2D video monitor~\cite{feng2020virtual}. This system also is triggered and controlled by voice commands.

AR systems for laparoscopy also use hand gestures and voice commands to control the actions used for the selection and manipulation of objects. In some works, the voice is combined with gaze direction (head gaze and eye gaze) as a trigger to control the draw and pointing functions. While voice commands can be confusing in the operation room, hand gestures demand that the surgeon pause the procedure to interact with the interface. Foot pedals, instead, are rather familiar for surgeons, as pedals are already used during laparoscopic procedures for cautery activation. Thus, the feet can be considered for interaction in augmented reality, such as using pedals to trigger and confirm visualization and navigation actions through the interface\cite{jayender2018novel}. Some adaptations allow the surgeon to perform many functions with the foot, such as slide through previous image exams in the interface panels, allied to the head-gaze for selecting and pointing \cite{zorzal2020laparoscopy}.

There are no guidelines, however, on how to apply head gaze, foot actions, and hand actions to control AR interfaces during surgery. 
We approach these issues in this paper, where we explore the design space and characterize the problems to be solved in an AR system for intraoperative assistance in laparoscopy.

\section{Materials and Methods}

Prior work shows that AR can provide several types of support for laparoscopic surgery. We are exploring interactions on AR interfaces to enhance the communication of the surgeon with local and remote stakeholders when needed and to remove the need for communication where it can benefit the operation outcome. To learn about these interactions, we propose a minimalist AR interface design.

To guide our design, we assumed a small set of requisites and constraints that are common to laparoscopic procedures:
\vspace{-1mm}
\begin{itemize}
\scriptsize
    \item the surgeon works standing;
    \vspace{-2mm}
    \item there is an inside-the-body video feed;
    \vspace{-2mm}
    \item preoperative images must be consulted intraoperatively;
    \vspace{-2mm}
    \item the surgeon wears sterile gloves;
    \vspace{-2mm}
    \item there are stakeholders in the OR and potentially remotely;
    \vspace{-2mm}
    \item the surgeon needs to indicate locations and objects in the surgical field to the stakeholders;
    \vspace{-2mm}
    \item AR headsets have a limited field of view;
\end{itemize}
    \vspace{-1mm}
Moreover, our design is aimed for usage as one of two components in a future mixed-reality application for remote collaboration in laparoscopic procedures. It means that there will be a second interface, not described here, where other people remotely located (students, experts) will follow the indications provided by the surgeon. Nevertheless, it can be used alone with the following setup and functionalities for within-OR collaboration. 

In the remainder of this section, we will present our UI design, the composition of the visualizations in the AR interface, the methods and hardware used to interact, and our proposed experimental setup for future user experiments. 

\subsection{Hardware and software platform}

We use Unity\footnote{https://unity.com/} 2019.4.20f1 LTS version for development, the latest version with support for Windows XR Plugin with \textit{Holographic Remoting}. This plugin provides an integration to connect and stream the application in the editor mode for the Microsoft® HoloLens 1st Generation\footnote{https://learn.microsoft.com/en-us/hololens/hololens1-hardware} (HL1) available in our lab. The HL1 has some limitations, and a major one is the 30 degrees field of view (FOV), which is smaller than the 54 degrees in the second-gen HL, and the average VR HMDs of about 90 degrees. This limitation causes the user to be unable to see virtual objects anywhere in their natural field of view. The holograms and interface elements appear only on a centered rectangle, requiring head turning to explore other areas. 
To simulate the live video from a laparoscopic source, we use a Logitech c525 webcam.

Inputs are needed for pointing locations and confirming options. We do not rely on special apparatus for pointing (see Sec.~\ref{ui_design} for details), but we had to adapt physical props to provide a fast and precise clicker for confirmations. We provide two options. One is pair of foot pedals similar to those already used in operating rooms. 

Although surgeons are familiar with pedals, additional pedals may be cumbersome. So, we provide a second option to allow hand interaction without having to release the surgical instruments. The device is an index finger ring with two 5mm push buttons on top of it, which the surgeon can wear under the sterilized gloves (see fig.~\ref{fig:teaser}b. The buttons are reachable with the thumb without taking the hand off the instrument. 

Both devices connect with the system through an Arduino Due plugged into a PC. The same software interprets the actions from either the pedals or buttons and executes them on the AR interface. The input functions for the pedals and ring buttons are described in Sec.~\ref{ui_design} below.

\subsection{Interface layout} \label{ui_design}

We identified in unstructured interviews with surgeons that they need to access the lap-video live feed from inside the body constantly, and they need to access preoperative data sporadically. Other information, such as vitals, is monitored by other members of the team. As virtual augmentations can be placed anywhere around the user, we prioritize displaying the lap-video and preoperative images in two adjacent panels viewed as holograms that the user can place wherever they wish around them in an egocentric perspective. Furthermore, the surgeon must be attentive to the real environment, the operation room, and the assistants during the procedure. Thus, our interface design avoids cluttering the view with unnecessary information and objects that could disturb concentration. Each of the two panels has specific functions, interactions, and virtual controllers with different actions. Panels can be seen in fig.~\ref{fig:teaser}a.

The most significant panel of the interface contains the real-time video from the laparoscopy procedure and its functionalities, initially positioned straight in front of the user's field of view. Right below the video panel is the image panel that compiles folders with preoperative exams, such as MRI, CT, and any planning sketches that could be helpful. Finally, in our experimental version, we included two lateral help panels containing usage indications for the interaction controls. For specific types of MIAS, more panels can be included, such as a vitals panel for cardiac procedures.




\subsection{Interaction design} \label{interaction_design}

The higher-level user tasks can be summarized as:
    \vspace{-2mm}

\begin{enumerate}
\scriptsize
    \item manipulate medical instruments;
    \vspace{-3mm}
    \item observe video feed;
    \vspace{-3mm}
    \item check for a specific preoperative image;
    \vspace{-3mm}
    \item signal a location in the video;
    \vspace{-3mm}
    \item circle an object in the video;
    \vspace{-3mm}
    \item save a screenshot;
    \vspace{-3mm}
    \item adjust the location of the hologram.
\end{enumerate}
    \vspace{-2mm}
Items 1 and 2 are inherent to the surgical task and are not supported by the interface. The other five tasks can be accomplished by a pointing plus confirmation metaphor with direct manipulation using the panels presented above.

While confirmation relies on either the ring or pedal buttons already described, we argue that, for pointing, head-gazing is an excellent solution. Notice that hands are busy and that the use of the tip of the surgical instruments for pointing will interfere with the actions significantly and will have limited scope. Few other options remain, such as foot pointing, voice, facial expressions, and BCI. None of them has reported great accuracy in previous works. Head-gaze pointing, in turn, has been extensively tried~\cite{Grinshpoon2018, trejos2015randomized, asao2021development}.

\subsubsection{Head-Gaze pointing}

The head gaze direction is captured from the HL1 orientation. The typical circular reticle pointer is shown where the gaze vector intercepts objects within the view, allowing for a mouse-pointer-like pointing. However, the reticle pointer provided by the head-gaze direction is sensible to tracking inaccuracies and user generated jittering. It can be challenging for most users to hit small targets, especially in the context of a narrow AR field of view. Thus, we propose to filter or scale down the mapping of the tracked motion to the reticle position. 

The average filtering approach stores the $n$ previous head-gaze directions and calculates a simple average. It renders the reticle at the respective averaged hit position. The scaling approach works by scaling down the head-reticle movement ratio that is $1$ when not scaled. On the disadvantages side, the average will add up latency while the scaling will decouple the center of the view from the target. If exaggerated, scaling will cause the user to turn too much the head that the target cannot be seen. As the display area follows the head, the target may even be cut out off the display area in excessive head turning.

\subsubsection{Selection and manipulation}

Each of the user tasks can be performed in the interface by pointing and clicking on elements that appear in the holographic panels described in sec.~\ref{ui_design}. In such a way, we designed each task as a combination of direct selection and manipulation actions, privileging efficiency. 

\begin{figure}[tb]
    \centering
    \includegraphics[width=0.9\linewidth]{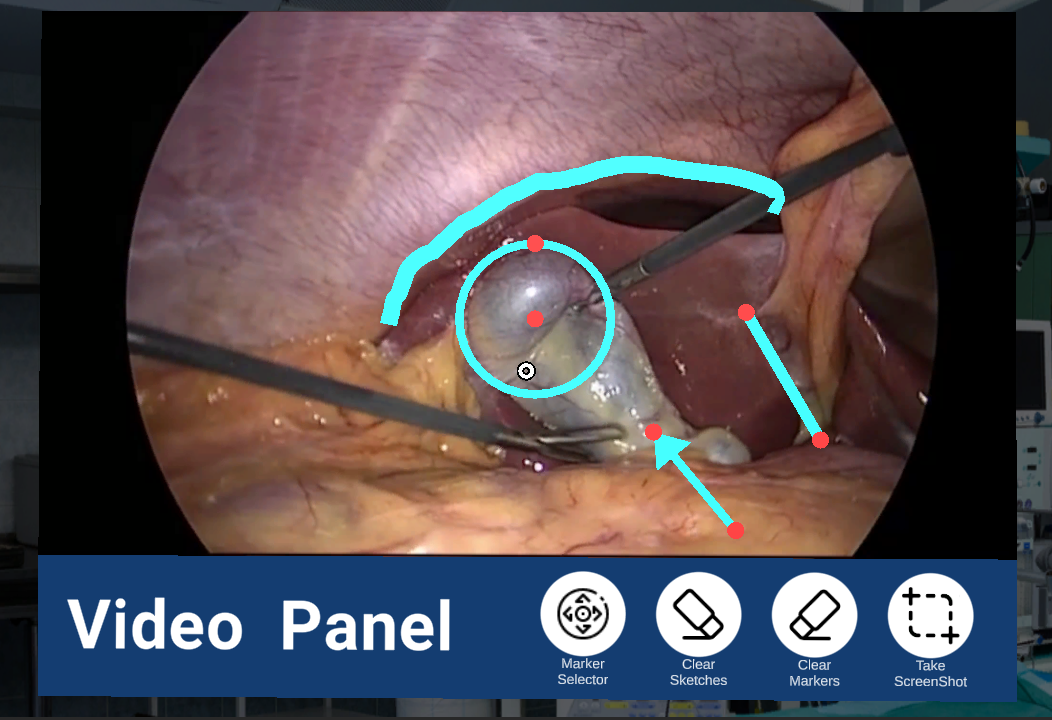}
    \caption{The video panel with annotations.}
    \label{fig:markers-video}
    \vspace{-4mm}
\end{figure}

In the live video panel, four actions can be initially triggered by selecting the respective button: place an annotation, clear markers, clear sketch, and screenshot (see fig.~\ref{fig:markers-video}). When $annotation$ is selected, a circular menu with four options allows choosing among line, arrow, circle, or free sketch. Line, arrow, and circle are considered markers. Markers can be deleted separately from the free sketch because surgeons often need to save markings, while the free sketch is used for incidental communication only. 

After choosing the desired marker, each has a specific set of placement/size actions. For $arrow$, the user first selects the arrowhead position and then moves to select the arrow tail location. For $line$, the process is similar. For $circle$, the user first selects the center location and then moves away to choose the radius. These markings can also be edited after creation. Red circular picking points appear at the line and arrow extremities and at the center and periphery of circles (see fig.~\ref{fig:markers-video}). Free sketching is applied by holding the button/pedal while moving. All selections described here are made with the left button/pedal.

Finally, the last possible action with the video panel involves all the panels in the interface. It supports the need of the user to modify the placement of the set of holograms to the most convenient location for each user and each moment of the operation. When the head-gaze reticle is in the video panel and the right button or pedal is pressed, the whole interface will follow the user's head-gaze to anywhere they wish until the right confirmation is activated, fixing the interface in the space. Besides setting up the most ergonomic location, this function also allows the surgeon to maintain the panels locked to the head like a heads-up display, which can be helpful to keep all the attention on the surgery, even when the head moves. 

\begin{figure}[h!]
    \centering
    \includegraphics[width=0.9\linewidth]{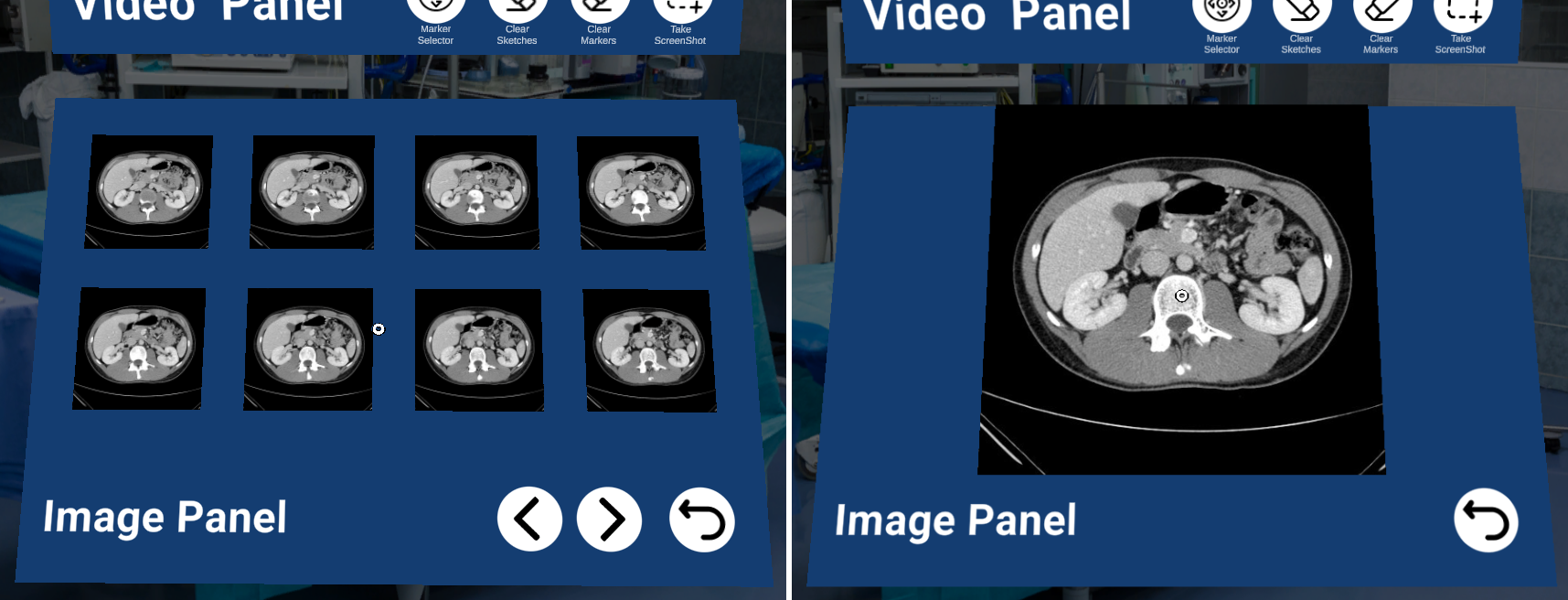}
    \caption{Left: the grid-view in the image panel. Right: Full view of a picture in the image panel.}
    \label{fig:image-panel}
        \vspace{-4mm}
\end{figure}

In the image panel (fig.~\ref{fig:image-panel}), the actions are more straightforward; the user can navigate through folders and the image grid by pointing the reticle to the virtual controllers and selecting the folders and the images by pressing on them. When an image is selected and presented in the panel size, the user can slide to the next or previous images using the right and left buttons respectively while the reticle is on the image. There is also a virtual back button to return to the image grid.

\subsection{User Experiment}

While we designed this interface accounting for the user's needs and general good practices of human-computer interaction, several aspects of the design need to be evaluated before they are final.

Among the aspects we wish to evaluate, there are: 
    \vspace{-2mm}

\begin{itemize}
\scriptsize
    \item the effects of averaging or scaling;
    \vspace{-2mm}
    \item how much scaling or averaging time is best;
    \vspace{-2mm}
    \item the effects of interacting with the foot or the hand;
    \vspace{-2mm}
    \item how much overhead the change of focus among panels imposes;
    \vspace{-2mm}
    \item how much communication improves with the use of the interface.
\end{itemize}
    \vspace{-2mm}
There is no room for a complete experimental evaluation in this workshop paper, but we overview in this section how we plan to characterize the affordances of intraoperative AR interfaces.

We split the research questions into three categories: range of parameters, surgical task, and communication efficiency. Therefore, we plan to conduct three experimental protocols with users. 

\begin{figure}[!tb]
    \centering
    \includegraphics[width=0.51\linewidth]{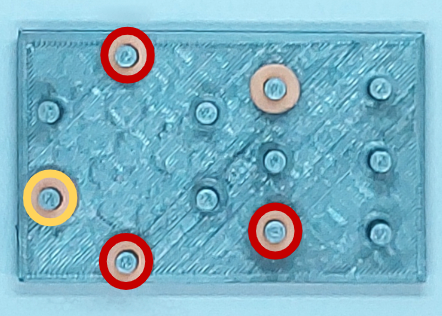}
    \includegraphics[width=0.47\linewidth]{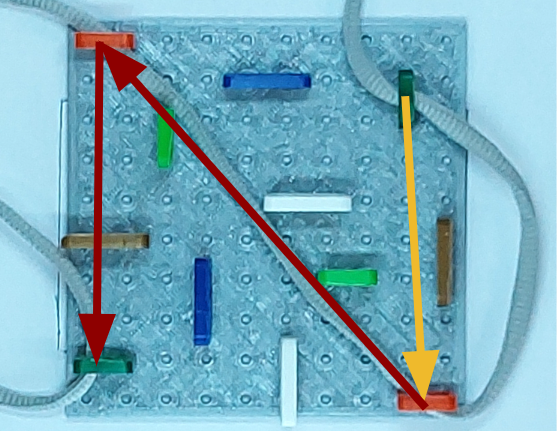}
    \caption{Task examples. Left: peg transfer. Right: Passing thread.}
    \label{fig:tasks}
    \vspace{-4mm}
\end{figure}

As it would be too premature to involve patients in this research, we developed two physical tasks similar to laparoscopy training using 3D-printed problems. The \textit{peg transfer task} (fig.~\ref{fig:tasks}a) is where the user transfers small rings using a couple of graspers from one peg to another in a peg board. The other task is thread passing (fig.~\ref{fig:tasks}b). The surgeon has to pass a thread through a sequence of holes standing on a board.

\subsubsection{Range of parameters}

We need to assess the effect of using the scaling or the averaging approach on user performance. We also need to define the most suitable level of scaling and averaging. Using two levels for each (low and high), we have four combinations. Adding the baseline with direct mapping, we obtain five interface conditions. Besides $interface$, we have another independent variable: $task$. This can be $peg$ or $thread$, but we wish to test problems of three different difficulty levels, which gives us six conditions. A $problem$ is a specific configuration of rings in pegs (more rings means more difficult) or a sequence of holes traversed by the thread (the more holes, the harder). A within-subjects design yields a collection of 30 unique combinations or samples per user. We should conduct a pilot test to check for feasibility/duration. Alternatively, if necessary, we can make $task$ between-subjects, which will require just 15 unique trials per user. 

For this test, we decided to remove the actual grasper manipulation in such a way that the users will only make annotations on the video. Thus, the population for this test does not need to be surgeons. We will ask participants to stand with their hands at the laparoscopy instruments, even though they are not performing the physical manipulation. The participants will consult the image panel for a problem, add the requested markings, save a screenshot and move to the next problem. As the hand is dominant for manipulation with the general population, this test will be conducted with the ring device only.

We will measure the precision of the annotations, the accuracy, and the time to complete each trial. We will then analyze for statistical relevance of the outputs regarding the different conditions.

\subsubsection{Surgical task}

Having decided from the previous test which parameters are best, they will be applied in the second test. Now, we will include the laparoscopic task and the foot pedal. The participants will be surgeons and apprentice surgeons at the university hospital. The test will be conducted using actual laparoscopic training white boxes.

The independent variables are the input device, the task, and the difficulty level. The participant will access the pictures in the image panel to study the problem and use the annotation tool while performing the laparoscopy training tasks. In the peg transfer task, the board will start with a different configuration of the pegs. The user will move to the configuration provided in the pictures and use the line marker to connect the origin and destination pegs. In the thread passing, the board will start with the thread passed in one role, and the user will pass through the others according to pictures in the image panel and annotate with a circle marker all the roles in which the thread passed.

We will measure the user's performance (precision, accuracy, time) and store segmented times for further evaluation of individual actions. We are interested in the variation of these measures by the input device and task type/level.

In all actions of the two experiments, we will also measure the head movement to analyze the effort overhead. 

\subsubsection{Communication efficiency}

It is too premature to present a protocol for communication efficiency. We will build it upon the results of the previous tests. We anticipate, however, that a second interface in immersive VR or desktop will be developed that a remote user will use in combination with the one presented here. Such an asymmetric collaborative approach may include broadcasting live 360 video from the operating room, and deictic annotations may be allowed from the remote participant as well.

\section{Conclusion and Future Work}

This exploratory work approaches affordances in AR interfaces for laparoscopy using the methodology of proposing a design and reporting the lessons learned. The work is not finished, but already raises some questions that are worth discussing with the community.

The current design carries the knowledge that the surgeons do not appreciate having two screens, one for live video and one paused for annotations, as we initially proposed. It also confirms the prior preference surgeons have for pedals in the OR, the need to check preoperative data on the fly, and that they rarely need to check vitals themselves. 

Moreover, the paper presents the planning for user experiments based on physical laparoscopy setups that will provide answers to some of the questions discussed. The AR interface proposed will also be part of a mixed-reality remote collaboration application in which a remote expert surgeon can guide a novice surgeon through virtual and augmented reality or a local professor can teach remote students. In this broader context, we also plan to include eye-tracking and investigate cognitive load, visual attention, and overall usability measures~\cite{Grandi2019}.

\acknowledgments{
This work is partly funded by the Brazilian agency \textit{Coordenação de Aperfeiçoamento de Pessoal de Nível Superior} (CAPES) - Finance Code 001, CNPq-Brazil project 311251/2020-0, by \textit{Fundação para a Ciência e Tecnologia} (Portuguese Foundation for Science and Technology) through grants 2022.09212.PTDC (XAVIER), UIDB/50021/2020, and Carnegie Mellon Portugal grant SFRH/BD/151465/2021 under the auspices of the UNESCO Chair on AI\&XR. We would like to thank Bytedance for their support with eye-tracking HMDs.
}

\bibliographystyle{abbrv-doi}
\nocite{jorge2011sketch,correia05,Pereira04,goncalves03,moreira11,kopper2008,johnson2013}
\bibliography{template}
\end{document}